%% file: paper.tex
\documentclass{libpaper/arxiv}
\pdfoutput=1  %
\usepackage{fullpage}
\usepackage{placeins}
\usepackage{hyperref}
\hypersetup{
    colorlinks,
    citecolor=black,
    filecolor=black,
    linkcolor=black,
    urlcolor=black
}

\input{libpaper/texheader}

\input{defs}
\input{macros}
\input{libpaper_metadata}

\input{libpaper/arxiv}

\begin{document}
\sloppy

\title{An Empirical Guide to the Behavior and Use of Scalable Persistent Memory}

\author{
  Jian Yang \and 
  Juno Kim \and 
	Morteza Hoseinzadeh \and
  Joseph Izraelevitz \and 
  Steven Swanson\footnote{Correspondence should be directed to swanson@cs.ucsd.edu.}
}
\date{
Nonvolatile Systems Laboratory\\
Computer Science \& Engineering\\
University of California, San Diego\\~\\
\today
}

\maketitle

\input{abstract}

\input{intro}

\input{background}
\input{basic}

\input{emulation}

\input{principle}

\input{limitations}

\input{related}
\input{conclude}
\input{bib}

\end{document}

%% file: libpaper/texheader.tex
\makeatletter                   %
\def\mdseries@tt{m}             %
\makeatother                    %

\usepackage[plain]{fancyref}
\usepackage[draft=true]{minted} %
\usepackage{color}
\usepackage{breakurl}           %

 \usepackage{lipsum}
\usepackage[final]{pdfpages}
\usepackage{setspace}
\usepackage{epsf}
\usepackage{epsfig}
\usepackage{graphicx}
\usepackage{algorithmic}
\usepackage{amsmath}
\usepackage{letltxmacro}

\usepackage{enumitem}
\usepackage{tikz}
\usepackage{libpaper/tightenum}

\usepackage{url}
\usepackage{multirow}

  \renewenvironment{thebibliography}[1]{%
    \begin{oldthebibliography}{#1}%
      \setlength{\parskip}{0ex}%
      \setlength{\itemsep}{0ex}%
  }%
  {%
    \end{oldthebibliography}%
  }

\input{libpaper/macros}

\input{libpaper/remark}

\input{libpaper/defs}

%% file: libpaper/macros.tex
\newwrite\arxivdeps
\immediate\openout\arxivdeps=\jobname-arxivdeps.log

\newcommand\verifymarkedforarxivfile[1]{%
\ifdefined\arxivbuild
\else
\IfFileExists{#1}%
{}%
{\GenericWarning{Marked file (#1) for inclusion in arxiv build doesn't exist}}%
\fi%
}

\newcommand\markforarxiv[1]{%
\verifymarkedforarxivfile{#1}%
\write\arxivdeps{IncludeInArxiv: #1}%
}

\DeclareUrlCommand\UScore{\urlstyle{rm}}

\LetLtxMacro\oldincludegraphics\includegraphics
\renewcommand{\includegraphics}[2][]{%
\markforarxiv{#2}%
\oldincludegraphics[#1]{#2}}

\LetLtxMacro\oldincludepdf\includepdf
\renewcommand{\includepdf}[2][]{%
\markforarxiv{#2}%
\oldincludepdf[#1]{#2}}

\def\nfigure[#1,#2,#3]{
\begin{figure}
\vspace*{0mm}
\begin{center}

\includegraphics[width=\columnwidth]{#1} 
\vspace*{-6mm}\caption[]{#2
} \label{#3}

\vspace*{-3mm}
\end{center}
\end{figure}}

\def\cfigure[#1,#2,#3]{
\begin{figure}
\vspace*{0mm}
\begin{center}

\includegraphics[width=3in]{#1} 
\vspace*{-3mm}\caption[]{#2
} \label{#3}
 
\vspace*{-5mm}
\end{center}
\end{figure}}

\def\cfigurefour[#1,#2,#3]{
\begin{figure}
\vspace*{0mm}
\begin{center}

\includegraphics[width=4in]{#1} 
\vspace*{-3mm}\caption[]{#2
} \label{#3}
 
\vspace*{-5mm}
\end{center}
\end{figure}}

\def\cfiguretemp[#1,#2,#3]{
\begin{figure}
\vspace*{0mm}
\begin{center}

\includegraphics[width=3.5in]{#1} 
\vspace*{-3mm}\caption[]{#2
} \label{#3}
 
\vspace*{-5mm}
\end{center}
\vspace*{-2mm}
\end{figure}}

\def\wfigure[#1,#2,#3]{
\begin{figure*}
\vspace*{0mm}
\begin{center}
 \includegraphics[width=\textwidth]{#1} 
 \vspace*{-3mm}\caption[]{#2
} \label{#3}
 
\end{center}
\end{figure*}}

\def\threefigure[#1,#2,#3,#4,#5]{
\begin{figure*}
\vspace*{0mm}
\begin{center}

\begin{tabular}{ccc}
\includegraphics[width=2in]{#1} & \includegraphics[width=2in]{#2} &  \includegraphics[width=2in]{#3} \\
(a) & (b) & (c) \\
\end{tabular}
\vspace*{-3mm}\caption[]{#4
} \label{#5}

\vspace*{-5mm}
\end{center}
\vspace*{-2mm}
\end{figure*}}

\def\dcfigure[#1,#2,#3,#4,#5,#6]{
{
\begin{figure*}
\begin{center}
\begin{minipage}[c]{\columnwidth}{
\includegraphics[width=\columnwidth]{#1} 
\vspace*{0mm}\caption[]{#2} \label{#3} \
}\end{minipage}\hspace*{\columnsep}\
\begin{minipage}[c]{\columnwidth}{
\includegraphics[width=\columnwidth]{#4} 
\vspace*{0mm}\caption[]{#5}\label{#6} \
}\end{minipage}
\end{center}
\end{figure*}
}
}

\def\scfigure[#1,#2,#3]{
{
\begin{figure*}
\begin{center}
\begin{minipage}[c]{3.5in}{
\includegraphics[width=3.5in]{#1} 
}\end{minipage}
\caption[]{#2} \label{#3} \
\end{center}
\end{figure*}
}
}

\def\tableByTable[#1,#2,#3,#4,#5,#6]{
{
\begin{table*}
\begin{center}
\begin{minipage}[c]{3in}{
\centering
{#1}
\vspace*{0mm}\tabcaption[]{#2}\label{#3} \
}\end{minipage}\hspace*{\columnsep}\
\begin{minipage}[c]{3in}{
\centering
{#4}
\vspace*{0mm}\tabcaption[]{#5}\label{#6} \
}\end{minipage}
\end{center}
\end{table*}
}
}

\def\figureByTable[#1,#2,#3,#4,#5,#6]{
{
\begin{figure*}
\begin{center}
\begin{minipage}[c]{3in}{
\centering
\includegraphics[width=\textwidth]{#1}
\vspace*{0mm}\figcaption[]{#2} \label{#3} \
}\end{minipage}\hspace*{\columnsep}\
\begin{minipage}[c]{3.3in}{
\centering
{#4}
\vspace*{0mm}\tabcaption[]{#5}\label{#6} \
}\end{minipage}
\end{center}
\end{figure*}
}
}

\def\tableByFigure[#1,#2,#3,#4,#5,#6]{
{
\begin{figure*}
\begin{center}
\begin{minipage}[c]{4.3in}{
\centering
{#1}
\vspace*{0mm}\tabcaption[]{#2} \label{#3} \
}\end{minipage}\hspace*{\columnsep}\
\begin{minipage}[c]{2.2in}{
\centering
\includegraphics[width=\textwidth]{#4}
\vspace*{-0.35in}\caption[]{#5}\label{#6} \
}\end{minipage}
\end{center}
\end{figure*}
}
}

\def\doublecfigure[#1,#2,#3,#4]{
{
\begin{figure}
\begin{center}
\begin{minipage}[c]{1.5in}{
\begin{center}
\includegraphics[width=1.5in]{#1}%
\end{center}
}\end{minipage}\hspace*{1em}\
\begin{minipage}[c]{1.5in}{
\begin{center}
\includegraphics[width=1.5in]{#2}%
\end{center}
}\end{minipage}
\vspace*{0mm}\caption[]{#3} \label{#4} \
\end{center}
\end{figure}
}
}

\def\qcfigure[#1,#2,#3,#4,#5,#6]{
{
\begin{figure*}
\vspace*{0.2in}\
\begin{center}
\begin{minipage}[c]{3in}{
\includegraphics[width=3in]{#1} 
\vspace*{-3mm}
}
\end{minipage}\hspace*{0.5in}\
\begin{minipage}[c]{3in}{
\includegraphics[width=3in]{#2} 
\vspace*{-3mm}
}\end{minipage}

\begin{minipage}[c]{3in}{
\includegraphics[width=3in]{#3} 
\vspace*{-3mm}
}
\end{minipage}\hspace*{0.5in}\
\begin{minipage}[c]{3in}{
\includegraphics[width=3in]{#4} 
\vspace*{-3mm}
}\end{minipage}
\end{center}
\caption[]{#5}\label{#6}
\end{figure*}
}
}

\def\twfigure[#1,#2,#3,#4,#5]{
{
\begin{figure*}
\vspace*{0.2in}\
\begin{center}
\begin{minipage}[c]{6.5in}{
\includegraphics[width=6.5in]{#1} 
\vspace*{-3mm}
}
\end{minipage}

\begin{minipage}[c]{6.5in}{
\includegraphics[width=6.5in]{#2} 
\vspace*{-3mm}
}\end{minipage}

\begin{minipage}[c]{6.5in}{
\includegraphics[width=6.5in]{#3} 
\vspace*{-3mm}
}
\end{minipage}
\end{center}
\caption[]{#4}\label{#5}
\end{figure*}
}
}

\def\dwfigure[#1,#2,#3,#4]{
{
\begin{figure*}
\vspace*{0.2in}\
\begin{center}
\begin{minipage}[c]{6.5in}{
\includegraphics[width=6.5in]{#1} 
\vspace*{-3mm}
}
\end{minipage}

\begin{minipage}[c]{6.5in}{
\includegraphics[width=6.5in]{#2} 
\vspace*{-3mm}
}\end{minipage}

\end{center}
\caption[]{#3}\label{#4}
\end{figure*}
}
}

\def\dssfigure[#1,#2,#3,#4,#5,#6]{
{
\begin{figure*}
\vspace*{0.2in}\
\begin{center}
\begin{minipage}[c]{4in}{
\includegraphics[width=4in]{#1}
\vspace*{-3mm}\caption[]{#2} \label{#3} \
}\end{minipage}\hspace*{0.5in}\
\begin{minipage}[c]{2in}{
\includegraphics[width=2in]{#4}
\vspace*{-3mm}\caption[]{#5}\label{#6} \
}\end{minipage}
\end{center}
\vspace*{-0.4in}\
\end{figure*}
}
}

\def\dsfigure[#1,#2,#3,#4,#5,#6]{
{
\begin{figure*}
\vspace*{0.2in}\
\begin{center}
\begin{minipage}[c]{3in}{
\includegraphics[width=3in]{#1}
\vspace*{-3mm}\caption[]{#2} \label{#3} \
}\end{minipage}\hspace*{0.5in}\
\begin{minipage}[c]{3in}{
\hspace*{0.5in}\
\includegraphics[height=3in]{#4}
\vspace*{-3mm}\caption[]{#5}\label{#6} \
}\end{minipage}
\end{center}
\vspace*{-0.4in}\
\end{figure*}
}
}

\def\dsyfigure[#1,#2,#3,#4,#5,#6]{
{
\begin{figure*}
\vspace*{0.2in}\
\begin{center}
\begin{minipage}[c]{2.5in}{
\includegraphics[height=2.5in]{#1}
\vspace*{-3mm}\caption[]{#2} \label{#3} \
}\end{minipage}\hspace*{0.5in}\
\begin{minipage}[c]{2.5in}{
\includegraphics[height=2.5in]{#4}
\vspace*{-3mm}\caption[]{#5}\label{#6} \
}\end{minipage}
\end{center}
\vspace*{-0.4in}\
\end{figure*}
}
}

\def\dyfigure[#1,#2,#3,#4,#5,#6]{
{
\begin{figure*}
\vspace*{0.2in}\
\begin{center}
\begin{minipage}[c]{3in}{
\includegraphics[height=3in]{#1} 
\vspace*{-3mm}\caption[]{#2} \label{#3} \
}\end{minipage}\hspace*{0.5in}\
\begin{minipage}[c]{3in}{
\includegraphics[height=3in]{#4} 
\vspace*{-3mm}\caption[]{#5}\label{#6} \
}\end{minipage}
\end{center}
\vspace*{-0.4in}\
\end{figure*}
}
}

\def\dyoldfigure[#1,#2,#3,#4,#5,#6]{
{
\begin{figure*}
\vspace*{0.2in}\
\begin{center}
\begin{minipage}[c]{3in}{
\epsfysize=2.0in\
\hspace{0.5in}\
\epsfbox{#1}
\vspace*{-3mm}\caption[]{#2} \label{#3} \
}\end{minipage}\hspace*{0.25in}\
\begin{minipage}[c]{3in}{
\epsfysize=2.0in\
\hspace{0.5in}\
\epsfbox{#4}
\vspace*{-3mm}\caption[]{#5}\label{#6} \
}\end{minipage}
\end{center}
\vspace*{-0.4in}\
\end{figure*}
}
}

\def\cfiguredouble[#1,#2,#3,#4]{
\begin{figure}
\vspace*{0.2in}\
\begin{center}
\begin{minipage}[c]{1.5in}{
\epsfxsize=1.5in\
\epsfbox{#1}
}\end{minipage}\hspace*{0.1in}\
\begin{minipage}[c]{1.5in}{
\epsfxsize=1.5in\
\vspace{0.1in}\epsfbox{#2}
}\end{minipage}\vspace*{-0.10in} \caption[]{#3}\label{#4}
\end{center}
\vspace*{-0.4in}\
\end{figure}
}

\def\wpfigure[#1,#2,#3,#4]{
\begin{figure*}
\vspace*{4mm}
\begin{center}

\includegraphics[width=#4]{#1} 

\vspace*{-3mm}\caption[]{#2
} \label{#3}

\vspace*{-5mm}
\end{center}
\end{figure*}}

\def\wprfigure[#1,#2,#3,#4,#5]{
\begin{figure*}
\vspace*{4mm}
\begin{center}

\includegraphics[width=#4, angle=#5]{#1} 

\vspace*{-3mm}\caption[]{#2
} \label{#3}

\vspace*{-5mm}
\end{center}
\end{figure*}}

\def\DoubleFigureWSlide[#1,#2,#3,#4,#5,#6,#7,#8,#9]{
\begin{figure*}
\vspace*{#9}
\begin{center}
\begin{minipage}{#4}
\includegraphics[width=#4]{#1}
\vspace*{-3mm}\caption{#2
}\label{#3}
\end{minipage}
\hspace{2em}
\begin{minipage}{#8}
\includegraphics[width=#8]{#5}
\vspace*{-3mm}\caption{#6
}\label{#7}
\end{minipage}
\vspace*{-5mm}
\end{center}
\end{figure*}
}

\def\DoubleFigureW[#1,#2,#3,#4,#5,#6,#7,#8]{
\begin{figure*}
\vspace*{0in}
\begin{center}
\begin{minipage}{#4}
\includegraphics[width=#4]{#1}
\vspace*{-3mm}\caption{#2
}\label{#3}
\end{minipage}
\hspace{2em}
\begin{minipage}{#8}
\includegraphics[width=#8]{#5}
\vspace*{-3mm}\caption{#6
}\label{#7}
\end{minipage}
\vspace*{-5mm}
\end{center}
\end{figure*}
}

\def\DoubleFigureWHack[#1,#2,#3,#4,#5,#6,#7,#8]{
\begin{figure*}
\vspace*{0in}
\begin{center}
\begin{minipage}{3in}
\includegraphics[width=#4]{#1}
\vspace*{-3mm}\caption{#2
}\label{#3}
\end{minipage}
\hspace{2em}
\begin{minipage}{3in}
\includegraphics[width=#8]{#5}
\vspace*{-3mm}\caption{#6
}\label{#7}
\end{minipage}
\vspace*{-5mm}
\end{center}
\end{figure*}
}

\def\ddcfigure[#1,#2,#3,#4]{
\begin{figure*}
\vspace*{0.2in}\
\begin{center}
\begin{minipage}[c]{\columnwidth}{
\includegraphics[width=\columnwidth]{#1} 
}\end{minipage}\hspace{0.5in}\
\begin{minipage}[c]{\columnwidth}{
\includegraphics[width=\columnwidth]{#2} 
}\end{minipage} \caption[]{#3}\label{#4}
\end{center}
\end{figure*}
}

\def\ddcfigureSlide[#1,#2,#3,#4,#5]{
\begin{figure*}
\vspace*{#5}\
\begin{center}
\begin{minipage}[c]{3in}{
\includegraphics[height=3in]{#1} 
}\end{minipage}\hspace{0.5in}\
\begin{minipage}[c]{3in}{
\includegraphics[height=3in]{#2} 
}\end{minipage}\vspace*{-0.10in} \caption[]{#3}\label{#4}
\end{center}
\vspace*{-0.4in}\
\end{figure*}
}

\def\cxfigure[#1,#2,#3]{
\begin{figure}
\vspace*{4mm}
\begin{center}
 
\epsfxsize=2.5in\
\epsfbox{#1}\
 
\vspace*{-0.10in}\caption[]{#2
} \label{#3}
 
\vspace*{-5mm}
\end{center}
\vspace*{-2mm}
\end{figure}}

%% file: libpaper/remark.tex
\newif\ifremark
\long\def\remark#1{
\ifremark%
        \begingroup%
        \dimen0=\columnwidth
        \advance\dimen0 by -1in%
        \setbox0=\hbox{\parbox[b]{\dimen0}{\protect\em #1}}
        \dimen1=\ht0\advance\dimen1 by 2pt%
        \dimen2=\dp0\advance\dimen2 by 2pt%
        \vskip 0.25pt%
        \hbox to \columnwidth{%
                \vrule height\dimen1 width 3pt depth\dimen2%
                \hss\copy0\hss%
                \vrule height\dimen1 width 3pt depth\dimen2%
        }%
        \endgroup%
\fi}

%% file: libpaper/defs.tex
\usepackage{xcolor}

\definecolor{cyanish}{rgb}{0,0.8,1.0}
\definecolor{orange}{rgb}{1.0,0.5,0.0}
\definecolor{pink}{rgb}{1.0,0.47,0.6}
\definecolor{light-gray}{gray}{0.95}
\definecolor{jiancolor}{RGB}{0,153,153}
\definecolor{mygreen}{RGB}{50,200,50}
\definecolor{pink}{rgb}{1.0,0.47,0.6}

\newcommand{\boldparagraph}[1]{\vspace*{1ex}\noindent\textbf{#1}\hspace{1em}}

\newcommand{\ignore}[1]{}

\newcommand{\myitem}[1]{\item \textbf{#1}}

\newcommand{\reffig}[1]{Figure~\ref{#1}}

\newcommand{\refsec}[1]{Section~\ref{#1}}

\newcommand{\us}{$\mu$s}
\newcommand{\x}[1]{$\times$}
\newcommand{\figtitle}[1]{\textbf{#1}}

%% file: defs.tex
\usepackage{booktabs}
\usepackage{subcaption}
\usepackage{tikz} 
\usepackage{pgfplots,pgfplotstable}
\usepackage{amsthm}
\usepgfplotslibrary{groupplots}

\usepackage{thmtools}
\usepackage{thm-restate}
\usepackage{etoolbox}

\makeatletter
\def\recommend@tions{}
\newcounter{recommend@cnt}

\newcommand{\allrecommendations}{\recommend@tions}
\makeatother

\newcommand{\takeaway}[2]{}

\definecolor{color-0}{rgb}{0.82,0.10,0.26}
\definecolor{color-1}{rgb}{0.00,0.00,1.00}
\definecolor{color-2}{rgb}{0.00,0.50,0.00}
\definecolor{color-3}{rgb}{1.00,0.65,0.00}
\definecolor{color-4}{rgb}{0.60,0.20,0.80}
\definecolor{color-5}{rgb}{0.57,0.51,0.32}
\definecolor{color-6}{rgb}{0.85,0.20,0.53}
\definecolor{color-7}{rgb}{0.33,0.33,0.33}
\definecolor{color-8}{rgb}{0.29,0.00,0.51}
\definecolor{color-9}{rgb}{0.00,0.29,0.29}

\newcommand{\embargo}[1]{Redacted}

\def\pcode{\texttt}

\newcommand{\XP}{3D\ XPoint}

\newcommand{\XPDIMM}{\XP{} DIMM}
\newcommand{\XPDIMMs}{\XP{} DIMMs}

\newcommand{\lattest}{LATTester}

\newcommand{\aepline}{XPLine}
\newcommand{\APC}{XPController}
\newcommand{\RMW}{XPBuffer}

\newcommand{\PM}{}

\newcommand{\PMLDRAM}{\PM{}DRAM}
\newcommand{\PMRDRAM}{\PM{}DRAM-Remote}
\newcommand{\PMLPMEM}{\PM{}Optane}
\newcommand{\PMLPMEMUI}{\PM{}Optane-NI}
\newcommand{\PMRPMEM}{\PM{}Optane-Remote}

\newcommand{\KernelVersion}{4.13.0}

\newcommand{\aepdimmsizegb}{256}
\newcommand{\aeptotalsizetb}{3}

\newcommand{\Ewrlong}{Effective Write Ratio}
\newcommand{\ewr}{EWR}
\newcommand{\ewrfull}{\Ewrlong (\ewr)}

\newcommand\seedatain[1]{}
\newcommand\seedatainas[2]{}

%% file: macros.tex
\def\ntwfigure[#1,#2,#3]{
\begin{figure*}
\vspace*{0mm}
\begin{center}
 \includegraphics[width=5in]{#1} 
 \markforarxiv{#1}
 \vspace*{-3mm}\caption[]{#2
} \label{#3}
 
\end{center}
\end{figure*}}

%% file: libpaper_metadata.tex
\newcommand{\gitcommit}{Uncommitted changes}
\def\arxivbuild{}

%% file: libpaper/arxiv.tex
\usepackage{fancyhdr}

%% file: abstract.tex
\begin{abstract}
After nearly a decade of anticipation, scalable nonvolatile memory
DIMMs are finally commercially available with the release of Intel's \XPDIMM{}. This new
nonvolatile DIMM supports byte-granularity accesses with access times on the
order of DRAM, while also providing data storage that survives power outages.

Researchers have not idly waited for real nonvolatile DIMMs (NVDIMMs)
to arrive.  Over the past decade, they have written a slew of papers proposing
new programming models, file systems, libraries, and applications built to
exploit the performance and flexibility that NVDIMMs promised to deliver.
Those papers drew conclusions and made design decisions without
detailed knowledge of how real NVDIMMs would behave 
or how industry would integrate them into computer
architectures.  
Now that \XP{} NVDIMMs are actually here, we can provide detailed performance 
numbers, concrete guidance
for programmers on these systems, reevaluate prior art for performance,
and reoptimize persistent memory software for the real \XPDIMM{}.

In this paper, we explore the performance properties and
characteristics of Intel's new \XPDIMM{} at the micro and macro
level. First, we investigate the basic characteristics of the device,
taking special note of the particular ways in which its performance
is peculiar relative to traditional DRAM or other past methods used
to emulate NVM.  From these observations,
we recommend a set of best practices to maximize
the performance of the device.	With our improved
understanding, we then explore the performance of prior art in application-level
software for persistent memory, taking note of where their performance was influenced
by our guidelines.
\end{abstract}

%% file: intro.tex
\section{Introduction}

Over the past ten years, researchers have been anticipating the arrival of
commercially available, scalable non-volatile main memory (NVMM) technologies
that provide ``byte-addressable'' storage that survives power outages. With the arrival
of Intel's Optane DC Persistent Memory Module (which we refer to as \XPDIMM{}s), we can
start to understand real capabilities, limitations, and characteristics of
these memories and starting designing and refining systems to fully leverage
them.

We have characterized the performance and behavior of
\XPDIMMs{} using a wide range of micro-benchmarks, benchmarks, and
applications.  The data we have collected demonstrate that many of the
assumptions that researchers have made about how NVDIMMs would behave and
perform are incorrect.

The widely expressed expectation was that NVDIMMs would have behavior that
was broadly similar to DRAM-based DIMMs but with lower performance (i.e.,
higher latency and lower bandwidth).  These assumptions are reflected in the
methodology that research studies used to emulate NVDIMMs, which include
specialized hardware platforms~\cite{pmfs}, software
emulation mechanisms~\cite{slottedpaging, quartz,nvheaps,strata, CDSMM},
exploiting NUMA effects~\cite{duan-date-2018,nvmfsbenchmarking, pvm}, and simply pretending DRAM
is persistent~\cite{dalinvm,atlas,makalu}.

We have found the actual behavior of \XPDIMM{}s to be more complicated and
nuanced than the ``slower, persistent DRAM'' label would suggest.  \XPDIMM{}
performance is much more strongly dependent on access size, access type (read
vs. write), pattern, and degree of concurrency than DRAM performance.  Furthermore,
\XPDIMM{}'s persistence, combined with the architectural support that Intel's
latest processors provide, leads to a wider range of design choices for software
designers.  

This paper presents a detailed evaluation of the behavior and performance of
\XPDIMMs{} on microbenchmarks and applications and provides concrete,
actionable guidelines for how programmers should tune their programs to make
the best use of these new memories.  We describe these guidelines, explore
their consequences, and demonstrate their utility by using them to guide the
optimization of several NVMM-aware software packages as well as the analysis of
previously-published results.

We also compare the behavior of real \XPDIMM{}s with several methods researchers
have used to emulate persistent main memory (e.g., using custom hardware,
exploiting NUMA effects, and pretending DRAM is persistent).  We find that all
of these emulation methodologies are inaccurate, suggesting that it is
unwise to assume that previously published results based on those
methodologies reflect performance on real hardware.

The paper proceeds as follows.  Section~\ref{sec:method} provides architectural
details on our test machine and the \XPDIMM{}.  Section~\ref{sec:basic} presents experiments
on microarchitectural details and parameters, and Section~\ref{sec:emulation}
provides a special focus on how the \XPDIMM{}
is different from DRAM and other emulation techniques.  
Section~\ref{sec:best_practice}
uses these results to posit best practices for programmers on the \XPDIMM{}.  In this section,
we first justify each guideline
with a micro-benchmark demonstrating the root cause.  We then present one or more case studies of
how the guideline would (and, in some cases, did)  influence a previously proposed \XP{}-aware software system.
Section~\ref{sec:limitations}
provides discussion as to how our guidelines extend to future generations of NVM.  Section~\ref{sec:related}
describes related work in this space, and Section~\ref{sec:conclude} concludes.

%% file: background.tex
\section{Background and Methodology}
\label{sec:method}

\wfigure[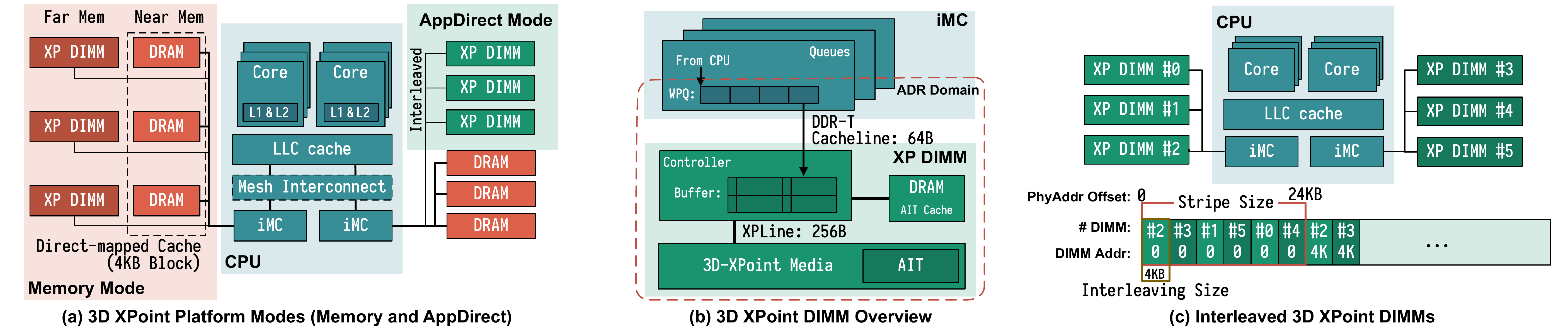, {\figtitle{Overview of (a) \XP{}
				platform, (b) \XP{} DIMM and (c) how \XP{} memories interleave across channels} \XP{} DIMM can work as
				volatile far memory with DRAM as cache or
				persistent memory for application accesses.},fig:overview]

In this section, we provide background on Intel's \XPDIMM{}, describe the test system, and then
describe the configurations we use throughout the rest of the paper.

\subsection{\XP{} Memory}
The \XPDIMM{}
is the first scalable, commercially available NVDIMM.
Compared to existing storage devices (including the
Optane SSDs) that connect to
an external interface such as PCIe, the \XPDIMM{} has lower latency, higher read bandwidth, and presents 
a memory address-based interface
instead of a block-based NVMe interface.
Compared to DRAM, it has higher density and 
persistence.
At its debut, the \XPDIMM{} is available in three different capacities: 128, 256, and 512~GB.

\subsubsection{Intel's \XPDIMM{}}

Like traditional DRAM DIMMs, the \XPDIMM{} sits on the memory bus, and connects to the processor's integrated memory controller~(iMC)
(Figure~\ref{fig:overview}(a)). Intel's Cascade Lake processors are the first (and only) CPUs to support \XPDIMM{}.
On this platform, each processor contains one or two processor dies which comprise separate NUMA nodes.
Each processor die has two iMCs, and each iMC supports three channels.
Therefore, in total, a processor die can support a total of six \XPDIMM{}s across its two iMCs.

To ensure persistence, the iMC sits within the \emph{asynchronous DRAM
refresh (ADR)} domain --- Intel's ADR feature ensures that CPU stores that
reach the ADR domain will survive a power failure (i.e.,\ will be flushed to 
the NVDIMM within the hold-up time, $< 100$~\us{})\cite{adr}.
The iMC maintains read and write pending queues (RPQs and WPQs) for each of the \XPDIMMs{} (\reffig{fig:overview}(b)),
and the ADR domain includes WPQs.  Once data reaches the WPQs, the ADR ensures that the iMC will flush the updates to \XP{} media on power failure. 
The ADR domain does not include the processor caches, so stores are only persistent once they reach the WPQs.

The iMC communicates with the \XPDIMM{} using the DDR-T interface in cache-line (64-byte) granularity.
This interface shares a mechanical and electrical interface with DDR4 but uses a different protocol
that allows for asynchronous command and data timing since \XP{} memory access latencies are not deterministic.

Memory accesses to the NVDIMM (\reffig{fig:overview}(b))
arrive first at the on-DIMM  controller (referred as \emph{\APC{}} in this paper), which
coordinates access to the \XP{} media.
Similar to SSDs, the \XPDIMM{} performs an internal address translation for wear-leveling and bad-block
management, and maintains an \emph{address indirection table} (AIT) for this translation~\cite{ait}.

After address translation, the actual access to storage media
occurs.  As the \XP{} physical media access granularity is 256 bytes (referred as \emph{\aepline{}} in this paper),
the \APC{} will translate smaller requests into larger 256-byte accesses, causing write amplification
as small stores become read-modify-write operations.
The \APC{} has a small write-combining buffer (referred as \emph{\RMW{}} in this paper), to merge adjacent writes.

It is important to note that all updates that reach the \RMW{} are already
persistent since \RMW{} resides within the ADR.  Consequently, the NVDIMM can buffer and
merge updates regardless of ordering requirements that the program specifies
with memory fences.

\subsubsection{Operation Modes}

\XPDIMM{}s can operate in two modes (\reffig{fig:overview}(a)):
Memory and App Direct.

\emph{Memory mode} uses \XP{} to expand main memory capacity without
persistence. It combines a \XPDIMM{} with a conventional DRAM DIMM on the same memory channel that
serves as a direct-mapped cache for the NVDIMM.  The cache block size is 64~B, 
and the CPU's memory controller manages the cache transparently.  The
CPU and operating system simply see the \XPDIMM{} as a larger (volatile) portion of main memory.

\emph{App Direct mode} provides persistence and does not use a DRAM cache.  The \XPDIMM{} appears
as a separate, persistent memory device.  The
system can install a file system or other management layer on the device to provide allocation, naming, and access to persistent data.  \XP{}-aware
applications and file systems can access the \XPDIMM{}s with load and store
instructions. 

In App Direct mode,  programmers can directly modify the \XPDIMM{}'s contents
using store instructions, and those stores will, eventually, become persistent. 
The cache hierarchy, however, can reorder stores, making recovery after a crash challenging~\cite{memorypersistency,justdologging,nvheaps,idologging,mnemosyne}.

Intel processors offer programmers a number of options to control store
ordering.  The instruction set provides \pcode{clflush} and \pcode{clflushopt}
instructions to flush cache lines back to memory, and \pcode{clwb} can write back
(but not evict) cache lines.
Alternately, software can use 
non-temporal stores (\pcode{ntstore}) to bypass the cache
hierarchy and write directly to memory.
All these instructions are non-blocking, so the program must issue an \pcode{sfence} to ensure that a previous cache flush, cache write back, or non-temporal store is complete and persistent.

In both modes, \XP{} memory can be (optionally) interleaved across channels
and DIMMs~(\reffig{fig:overview}(c)). On our platform, the only supported interleaving size is 4~KB, which ensures that accesses to
a single page fall into a single DIMM.  With six DIMMs, an access larger than 24~KB will access all the DIMMs.

\subsection{System Description}

We perform our experiments on a dual-socket evaluation platform
provided by Intel Corporation.  
The CPUs are 24-core Cascade Lake engineering samples with the similar spec as the previous-generation Xeon Platinum 8160.
Each CPU has two iMCs and six memory channels (three channels per iMC). A 32~GB Micron DDR4
DIMM and a \aepdimmsizegb{}~GB Intel \XPDIMM{}  attach to each
of the memory channels.  Thus the system has 384~GB (2 socket \x{} 6 channel \x{}
32~GB/DIMM) of DRAM, and \aeptotalsizetb{}~TB (2 socket \x{} 6 channel \x{}
\aepdimmsizegb{}~GB/DIMM) of NVMM.
Our machine runs Fedora 27 with Linux kernel version
\KernelVersion{} built from source.

\subsection{Experimental Configurations}

As the \XPDIMM{} is both persistent and byte-addressable, it can fill the role
of either a main memory device (i.e., replacing DRAM) or a persistent device
(i.e., replacing disk). In this paper, we focus on the persistent usage,
and discuss how our findings apply to using \XPDIMM{} as volatile memory in~\refsec{sec:limitations}.

Linux manages persistent memory by creating \pcode{pmem} namespaces over a contiguous span of physical memory.
A namepace can be backed by interleaved or non-interleaved \XP{} memory, or emulated persistent
memory backed by DRAM.
In this study, we configure the \XP{} memory in App Direct mode and create a namespace for
each type of memory.

Our baseline (referred as \emph{\PMLPMEM{}}) exposes six \XPDIMM{}s from the same socket as a single interleaved namespace.
In our experiments, we use local accesses (i.e., from the same NUMA node) as the baseline to compare 
with one or more other configurations, such as
access to \XP{} memory on the remote socket (\emph{\PMRPMEM{}}) or DRAM on the local or remote socket (\emph{\PMLDRAM{}} and \emph{\PMRDRAM{}}). 
To better understand the raw performance of \XP{} memory without interleaving, we also
create a namespace consisting of a single \XPDIMM{} and denote it as \emph{\PMLPMEMUI{}}.

%% file: basic.tex
\section{Performance Characterization}
\label{sec:basic}

In this section, we measure \XP{}'s performance along multiple axes to provide
the intuition and data that programmers and system designers will need to
effectively utilize \XP{}.  We find that \XP{}'s performance characteristics
are complex and surprising in many ways, especially relative the notion that
persistent memory behaves like slightly-slower DRAM.

\subsection{\lattest{}}

Characterizing \XP{} memory is challenging for two reasons.
First, the underlying technology has major differences from DRAM
but publicly-available documentation is scarce.
Secondly, existing tools measure memory performance primarily as a function of locality and access size,
but we have found that \XP{} performance depends strongly on memory interleaving and concurrency as well.
Persistence adds an additional layer of complexity for performance measurements.

To fully understand the behavior of the \XP{} memory, we built a microbenchmark toolkit
called \lattest{}.  To accurately measure the CPU cycle count and minimize the impact from
the virtual memory system, \lattest{} runs as a dummy file
system in the kernel and accesses pre-populated (i.e., no page-faults) kernel virtual addresses of
\XP{} DIMMs.  \lattest{} also pins the kernel threads to fixed
CPU cores and disables IRQ and cache prefetcher.

In addition to simple latency and bandwidth measurements, \lattest{} collects a
large set of hardware counters at both the CPU and NVDIMM.

Our investigation of \XP{} memory behavior proceeds in two phases.  First, we
perform a broad, systematic  ``sweep'' over \XP{} configuration parameters including access
patterns (random vs. sequential), operations (loads, stores, fences, etc.),
access size, stride size, power budget, NUMA configuration, and address space interleaving.

Then, we design targeted experiments to investigate anomalies in that data and verify or disprove our hypotheses
about the underlying cause.
Between our initial sweep and the follow-up tests, 
we collected over ten thousand data points.

\subsection{Typical Latency}
\label{sec:idle-lat}

Read and write latencies are key memory technology parameters. We measure read latency by timing the average latency for
individual 8-byte load instructions to sequential and random memory addresses.
To eliminate caching and queuing effects, we empty the CPU pipeline and issue a memory
fence (\pcode{mfence}) between measurements (\pcode{mfence} serves the purpose of serialization for reading timestamps).
For writes, we load the cache line into the cache and then measure the latency of one of two instruction sequences: a 64-bit store, a \pcode{clwb}, and an \pcode{mfence}; or
a non-temporal store followed by an \pcode{mfence}.

These measurements reflect the load and store latency as seen by software
rather than those of these underlying memory device.  For loads, the latency
includes the delay from the on-chip interconnect, iMC, \APC{} and the actual \XP{} media.
Our results (\reffig{fig:raw_latency}) show the read latency for \XP{} is 2\x{}--3\x{} higher than DRAM.
We believe most of this difference is due to \XP{} having a longer media latency.
\XP{} memory is also more
pattern-dependent than DRAM.  The random-vs-sequential gap is 20\% for DRAM but 80\% for \XP{} memory,
and we believe this gap is a consequence of the \RMW{}.  
For stores, the memory store and fence instructions commit once the data reaches the ADR at the iMC. Both DRAM and \XP{} memory
show a similar latency. Non-temporal stores are more expensive than writes with cache flushes (\pcode{clwb}).

In general, the latency variance
for \XP{} is extremely small, save for an extremely small number of ``outliers'', which we investigate in the next section.
The sequential read latencies for \XPDIMM{}s have higher variances, as the first cache line access loads the entire \aepline{}
into \RMW{}, and the following three accesses read data in the buffer.

\ntwfigure[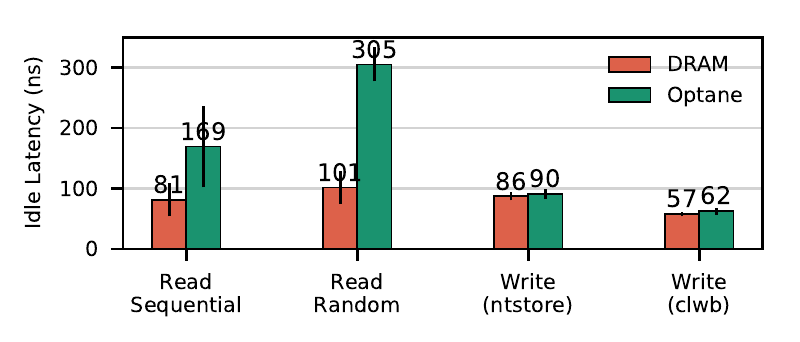,{\figtitle{Best-case latency} An
			experiment showing random and sequential read latency, as well as write
			latency using cached write with \pcode{clwb} and \pcode{ntstore} instructions.  Error bars show
			one standard deviation.},fig:raw_latency]

\subsection{Tail Latency}

Memory and storage system tail latency critically affects response times and
worst-case behavior in many systems. In our tests, we observed a very consistent
latency for loads and stores except a few ``outliers'', which
increase in number for stores when accesses are concentrated in a ``hot spot''.

\reffig{fig:tail_latency} measures the relationship between tail latency and
access locality.  The graph shows the 99.9th, 99.99th, and maximal latencies
as a function of hot spot size. The number of outliers (especially for the ones over 50\us{})
reduces as the hotspot size increases and do not exist for DRAM.

These spikes are rare (0.006\% of the accesses), but their latency
are 2 orders of magnitude higher than a common case \XP{} access.
We suspect this effect is due to remapping for wear-leveling or thermal concerns, but we cannot be sure.

\ntwfigure[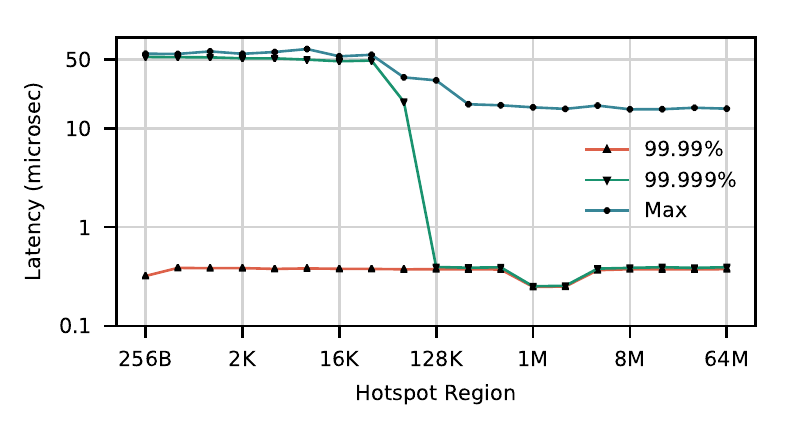,{\figtitle{Tail latency} An experiment
			showing the tail latency of writing to a small area of memory (hotspot)
			sequentially. \XP{} memory has rare `outliers' where a small number of writes take
			up to 50~\us{} to complete (an increase of 100\x{} over the usual latency).},fig:tail_latency]

\subsection{Bandwidth}
\label{sec:max-bw}

Detailed bandwidth measurements are useful to application designers as 
they provide insight into how a memory technology
will impact overall system throughput. 
First, we measure \XP{} and DRAM bandwidth for random and sequential reads and writes
under different levels of concurrency.  

\reffig{fig:bwthreads} shows the
bandwidth achieved at different thread counts for sequential accesses with 256~B access granularity.
We show loads and stores (\texttt{Write(ntstore)}),
as well as cached writes with flushes (\texttt{Write(clwb)}).  All experiments use AVX-512 instructions and access
the data at 64~B granularity.  The left-most graph plots performance for interleaved DRAM accesses,
while the center and right-most graphs plot performance for non-interleaved and interleaved \XP{}. In the non-interleaved measurements all the accesses go to a single DIMM.

\reffig{fig:bw} shows how performance varies with access size.  The
graphs plot aggregate bandwidth for random accesses of a given size.  We use
the best-performing thread count for each curve (given as ``<load thread
count>/<ntstore thread count>/<store+clwb thread count>'' in the figure).
Note that the best performing thread count for \PMLPMEM{}(Read) varies with different access sizes for random accesses,
where 16 threads show good performance consistently.

The data shows that DRAM bandwidth is both significantly higher than \XP{}
and scales predictably (and monotonically) with thread count until
it saturates the DRAM's bandwidth and that bandwidth is mostly independent of
access size.  

The results for \XP{} are wildly different. First, for a single DIMM, the maximal read bandwidth
is 2.9\x{} of the maximal write bandwidth (6.6~GB/s and 2.3~GB/s respectively), where DRAM has a smaller gap (1.3\x{}) between read and write bandwidth. 

Second, with the exception of interleaved reads, \XP{} performance is non-monotonic
with increasing thread count.  For the non-interleaved (i.e., single-DIMM)
cases, performance peaks at between one and four threads and then tails off.
Interleaving pushes the peak to twelve threads for \pcode{store+clwb}.
We will return to the negative impact of rising thread count on performance in \refsec{sec:ewr}.

Third, \XP{} bandwidth for random accesses under 256~B is poor.  This ``knee'' corresponds to \aepline{} size.
DRAM bandwidth does not exhibit a similar ``knee''
at 8~kB (the typical DRAM page size), because the cost of opening a page of
DRAM is much lower than accessing a new page of \XP{}.

Interleaving (which spreads accesses across all six DIMMs) adds further complexity: 
\reffig{fig:bwthreads}(right)
and \reffig{fig:bw}(right) measure bandwidth across six interleaved NVDIMMs.
Interleaving improves
peak read and write bandwidth by 5.8\x{} and 5.6\x{}, respectively.
These speedups match the number of DIMMs in the system and highlight the
per-DIMM bandwidth limitations of \XP{}. The most striking feature of the graph is
a dip in performance at 4~KB --- this dip is an emergent effect caused by contention at the iMC,
and it is maximized when threads perform random accesses close to the interleaving size. 
We will further discuss the cause and solution of this issue in \refsec{sec:interleaving}.

\wfigure[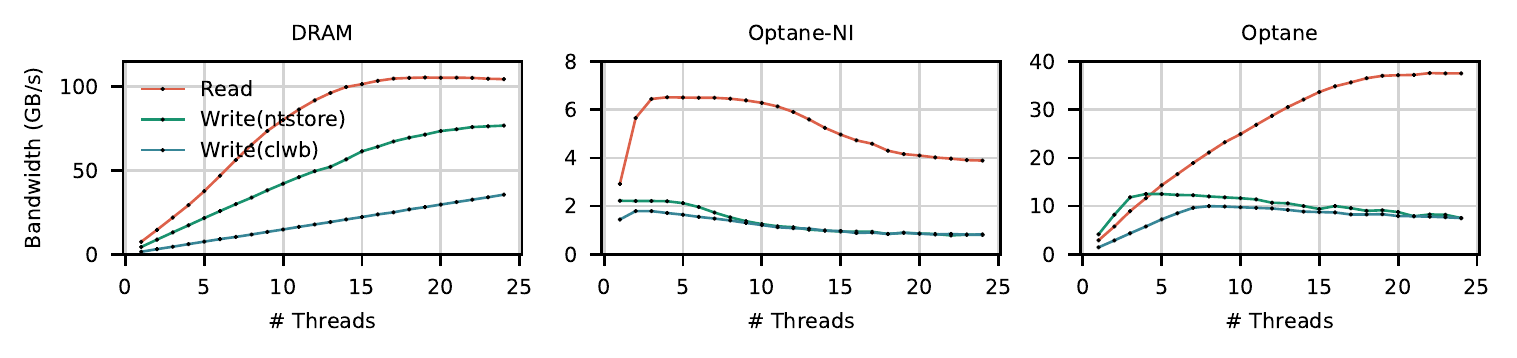,{\figtitle{Bandwidth vs. thread count} An experiment showing
maximal bandwidth as thread count increases (from left to right) on local DRAM, non-interleaved
and interleaved \XP{} memory. All threads use a 256~B access size. (Note the difference in vertical scales). \vspace{-8mm}} ,fig:bwthreads]
	
\wfigure[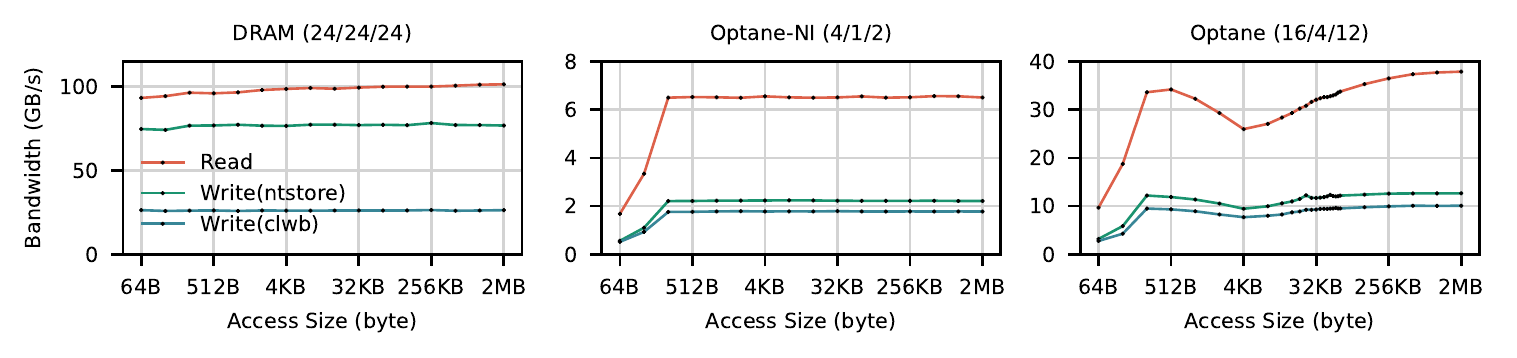,{\figtitle{Bandwidth over access size} An experiment showing
    maximal bandwidth over different access sizes on (from left to right) local DRAM, interleaved
    and non-interleaved \XP{} memory. Graph titles include the number of threads used
    in each experiment (\pcode{Read/Write(ntstore)/Write(clwb)}). \vspace{-8mm}} ,fig:bw]

\subsection{Latency Under Load}

In most systems, latency increases slowly as bandwidth rises until it reaches a
``wall'' and queuing effects cause latency to rise sharply.  To explore this
effect in \XP{} and DRAM, we use \lattest{} to vary the offered load to each
type of memory.  We use the thread counts that get consistently good performance across access sizes.
For loads, we use 16 threads.  For non-temporal stores, we use
4 threads. 
Each thread performs cache line sized memory accesses
and delays for a set interval between two accesses.
\reffig{fig:loaded_latency} plots the results as we vary the delay interval
from 0 to 80~\us{}.

The data shows that both DRAM and \XP{} behave as expected. The
``wall'' occurs earlier for \XP{} and the latencies are significantly
higher. \XP{} also shows an imbalance between sequential and random access
patterns, unlike DRAM. 

\ntwfigure[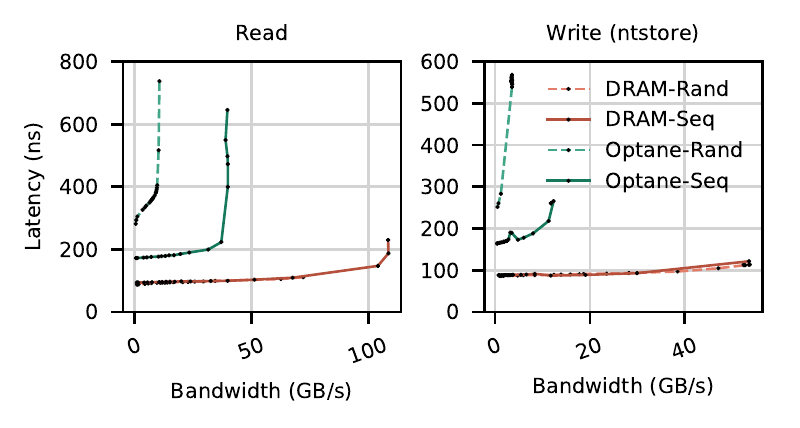,{\figtitle{Memory latency and
				bandwidth under varying load}
			The ``knee'' in the graph occurs when the device begins
			to suffer from queuing effects and maximum bandwidth is reached. DRAM memory
			performance is consistent between sequential and random accesses, while \XP{}
			memory is much more sensitive to the access pattern.
		},fig:loaded_latency]

%% file: emulation.tex
\section{Comparison to Emulation}
\label{sec:emulation}

Non-volatile memory research has been popular in recent years
(e.g.~\cite{dalinvm, mnemosyne, atlas, novapaper, orion, memorypersistency, pmorder, strata, mojim, CDSMM, SoupFS, espresso, slottedpaging, justdologging}).
However, since scalable NVDIMMs have not been available,
most of the NVM based systems have been evaluated on emulated
NVM.  Common ways to emulate NVM include adding delays to memory accesses in software~\cite{mnemosyne, strata, CDSMM}, using software emulators~\cite{slottedpaging, quartz},
using software simulation~\cite{memorypersistency, pmorder,nvheaps},
using hardware emulators such as Intel's Persistent Memory Emulator Platform
(PMEP)~\cite{pmfs} to limit latency and bandwidth~\cite{mojim, nova-fortis, novapaper}, 
using DRAM on a remote socket (\PMRDRAM{})~\cite{nvmfsbenchmarking, pvm,duan-date-2018}, underclocking DRAM~\cite{justdologging}
or just using plain DRAM~\cite{dalinvm, atlas, orion, octopus} or battery-backed DRAM~\cite{SoupFS}.

Below, we compare these emulation techniques to real \XP{} using
microbenchmarks and then provide a case study in how those differences can affect
research results.

\subsection{Microbenchmarks in Emulation}

\reffig{fig:microemu}(left) shows the write latency/bandwidth curves
for NVM emulation mechanisms (e.g.\ PMEP, \PMRDRAM{}, DRAM)
in comparison to real \XP{} memory
(similar to Figure~\ref{fig:loaded_latency}).
\reffig{fig:microemu}(right) shows bandwidth with respect
to the number of reader/writer threads (all experiments
use a fixed number of threads that give maximum bandwidth).
Our PMEP configuration adds a
300~ns latency on load instructions and throttles write bandwidth at 1/8 of
DRAM bandwidth as this configuration is the standard used in previous works~\cite{novapaper,nova-fortis,mojim,NVMMAPP-study}.
Note that PMEP is a specialized hardware platform, so its
performance numbers are not directly comparable to the system we use in our other
experiments.  

The data in these figures shows that none of the emulation mechanism captures the details of
\XP{}'s behavior --- all methods deviate drastically in their performance from real \XP{} memory.
They fail to capture \XP{} memory's preference for sequential accesses and read/write asymmetry
and give wildly inaccurate guesses for device latency and bandwidth.

\ntwfigure[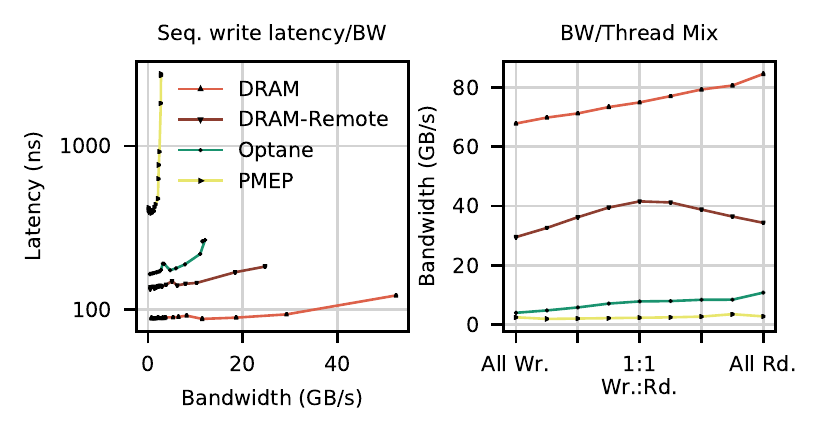,{\figtitle{Microbenchmarks under emulation}
    The emulation mechanisms used to evaluate many projects do not accurately
    capture the complexities of \XP{} performance.},fig:microemu]

\subsection{Case Study: Optimizing RocksDB}

The lack of emulation fidelity can have a dramatic effect on the performance of software.
Results may be misleading, especially
those based on simple DRAM.
In this section, we revisit prior art in NVM programming in order
to demonstrate that emulation is generally insufficient to capture the performance
of \XPDIMM{}s and that future work should be validated on real hardware.

\ntwfigure[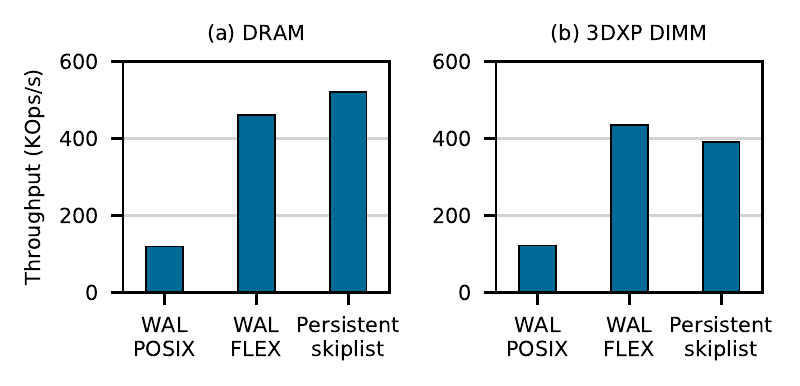,{\figtitle{Migrating RocksDB to \XP{} Memory} 
\XP{} media is sufficiently different from DRAM to invert prior conclusions.
Using a persistent memtable works best for 
DRAM emulating persistent memory, but 
on real \XP{} memory the conclusion is reversed.},fig:rocksdb-emu]

RocksDB~\cite{rocksdb} is a high-performance embedded key-value store, designed by Facebook and inspired
by Google's LevelDB~\cite{leveldb}. RocksDB's design is centered around
the log-structured merge tree (LSM-tree),
which is designed for block-based storage devices, absorbing random writes and converting them to
sequential writes to maximize hard disk bandwidth.

A recent study~\cite{flex} compared two strategies for adapting RocksDB to use
persistent memory. The first uses a fine-grained persistence approach to
migrate RockDB's ``memtable'' to persistent memory, eliminating the need for a
file-based write-ahead log.  The second approach moved the write-ahead log to
persistent memory and used a simpler acceleration technique called FLEX to move
logging performance.

The study used DRAM as a stand-in for \XP{}, and found that fine-grained
persistence offered 19\% better performance.

We replicated these experiments
on real \XP{} media.  We used the RocksDB test \texttt{db\_bench} on
SET throughput with 
20-byte key size and 100-byte value size, and sync the database after each SET operation; 
the results are shown in Figure~\ref{fig:rocksdb-emu}.
With real \XP{}, the result is the opposite: FLEX performs better than fine-grained persistence by 10\%.
These results are not surprising given \XP{} memory's preference for sequential accesses
and its problem with small random writes.

\subsection{Discussion}

We make two broader observations about our results.  First, the differences
between emulated and real persistent memory are large enough to alter the
conclusions that researchers might draw from their experiments.  Second, there
does not appear to be any simple relationship between emulated results and
results on real \XP{} hardware.

We conclude that basing future designs (and design decisions) on the results of
emulation-based studies may cause system designers to overlook superior
options.  It also demonstrates the value of re-evaluating previously considered
(and perhaps discarded) ideas on new hardware.

%% file: principle.tex
\section{Best Practices for \XPDIMM{}s}
\label{sec:best_practice}

\refsec{sec:basic} highlights the many differences between \XP{} and conventional storage and memory technologies,
and \refsec{sec:emulation} shows how these differences can manifest to invalidate macro-level results.  These differences mean that existing intuitions about how to optimize software for disks and memory do not apply directly to \XP{}.   This section distills the results of our
characterization experiments into a set of four principles for how to build and tune \XP{}-based systems.

\begin{samepage}
\begin{enumerate}
  \myitem{Avoid random accesses smaller than $<$ 256~B.}
  \myitem{Use non-temporal stores when possible for large transfers, and control of cache evictions.}
	\myitem{Limit the number of concurrent threads accessing a \XP{} DIMM.}
  \myitem{Avoid NUMA accesses (especially read-modify-write sequences)}.
\end{enumerate}
\end{samepage}

Below, we describe the guidelines in detail, give examples of how to implement them, and provide case studies in their application.

\input{guideline-seq}

\input{guideline-ntstore}

\input{guideline-par}
\input{guideline-numa}

%% file: guideline-seq.tex
\subsection{Avoid small random accesses}
\label{sec:ewr}

Internally, \XP{} DIMMs update \XP{} contents at a 256~B granularity.  This
granularity, combined with a large internal store latency, means that smaller
updates are inefficient since they require the DIMM to perform an internal
read-modify-write operation causing write amplification.  The less locality the accesses exhibit, the more
severe the performance impact.

To characterize the impact of small stores, we perform two experiments.
First, we quantify the inefficiency of small stores using a metric we have
found useful in our study of \XP{}.   The \emph{\ewrfull{}} is the ratio of bytes issued by the iMC divided by the
number of bytes actually written to the \XP{} media (as measured by the DIMM's hardware
counters).  \ewr{} is the inverse of write amplification.

\ewr{} values below one indicate the \XPDIMM{} is operating inefficiently
since it is writing more data internally than the application requested.  The
\ewr{} can be greater than one, due to write-combining at the \RMW{} or
DRAM caching (when the \XP{} is operating in Memory Mode).

\reffig{fig:ewr} plots the strong correlation between \ewr{} and effective device
bandwidth for a single DIMM for all the measurements in our systematic sweep of
\XP{} performance.  Based on this relationship, we conclude that working to maximize
\ewr{} is a good way to maximize bandwidth.

In general, small stores exhibit \ewr{}'s of less than one.  For example, when
using a single thread to perform non-temporal stores to random
accesses, it achieves an \ewr{} of 0.25 for 64-byte accesses and 0.98 for
256-byte accesses.

\ntwfigure[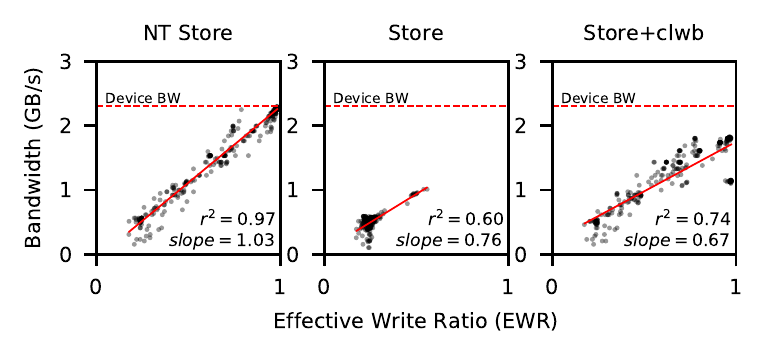,{\figtitle{Relationship between EWR and
				throughput on a single DIMM} Each dot represents
			an experiment with different access size, thread count
			and power budget configurations. Note the correlation between the metrics.},fig:ewr]

Notably, 256-byte updates are efficient, even though the iMC only issues 64~B
to accesses the DIMM --- the \RMW{} is responsible for buffering and
combining 64~B accesses into 256~B internal writes.  As a consequence,
\XP{} DIMMs can efficiently handle small stores, if they exhibit sufficient locality.
To understand how much locality is sufficient, we crafted an experiment to
measure the size of the \RMW{}.  First, we allocate a contiguous region of $N$
\aepline{}s.  During each ``round'' of the experiment, we first update the
first half (i.e., 128~B) of each \aepline{} in turn.  Then, we update the
second half of each \aepline{}.  We measure the \ewr{} for each
round. \reffig{fig:buffers} shows the results.

Below $N=64$ (that is, a region size of 16~KB), the \ewr{} is near unity, suggesting that the accesses to the
second halves are hitting in the \RMW{}.  Above $N=64$, write amplification
jumps, indicating a sharp rise in the miss rate.  This result implies the
\RMW{} is approximately 16~KB in size. Further experiments demonstrate that
reads also compete for space in the \RMW{}.

Together these results provide specific guidance for maximizing \XP{} store
efficiency: Avoid small stores, but if that is not possible, limit the working
set to 16~KB per \XP{} DIMM.

\ntwfigure[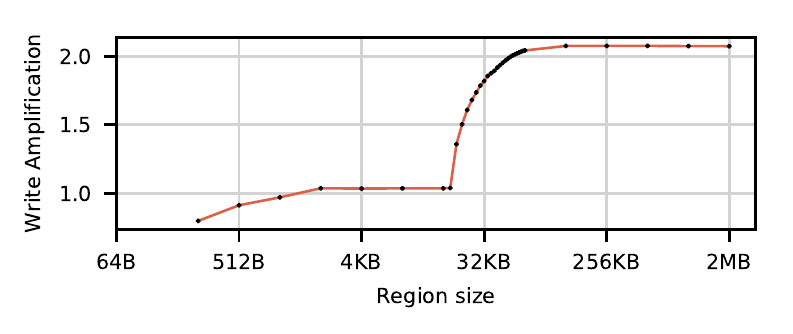,{\figtitle{Inferring \RMW{} capacity}
    The data show that the \XP{} DIMM can use the \RMW{} to coalesce writes spread across up to 64 \aepline{}s.},fig:buffers]

\subsubsection{Case Study: RocksDB}

The correlation between \ewr{} bandwidth explains the results for RocksDB seen
in \refsec{sec:emulation} and \reffig{fig:rocksdb-emu}.  The persistent
memtable resulted in many small stores with the poor locality, leading to a low
\ewr{} of 0.434.  In contrast, the FLEX-based optimization of WAL uses
sequential (and larger) stores, resulting in an \ewr{} of 0.999.

\subsubsection{Case Study: The NOVA filesystem}

\input{nova-embed}

%% file: nova-embed.tex
NOVA~\cite{novapaper, nova-fortis} is a log-structured, NVMM file system
that maintains a separate log for each file and directory and uses copy-on-write
for file data updates to ensure data consistency.  The original NOVA studies used emulated NVMM for their
evaluations, so NOVA has not been tuned for \XP{}.

\ntwfigure[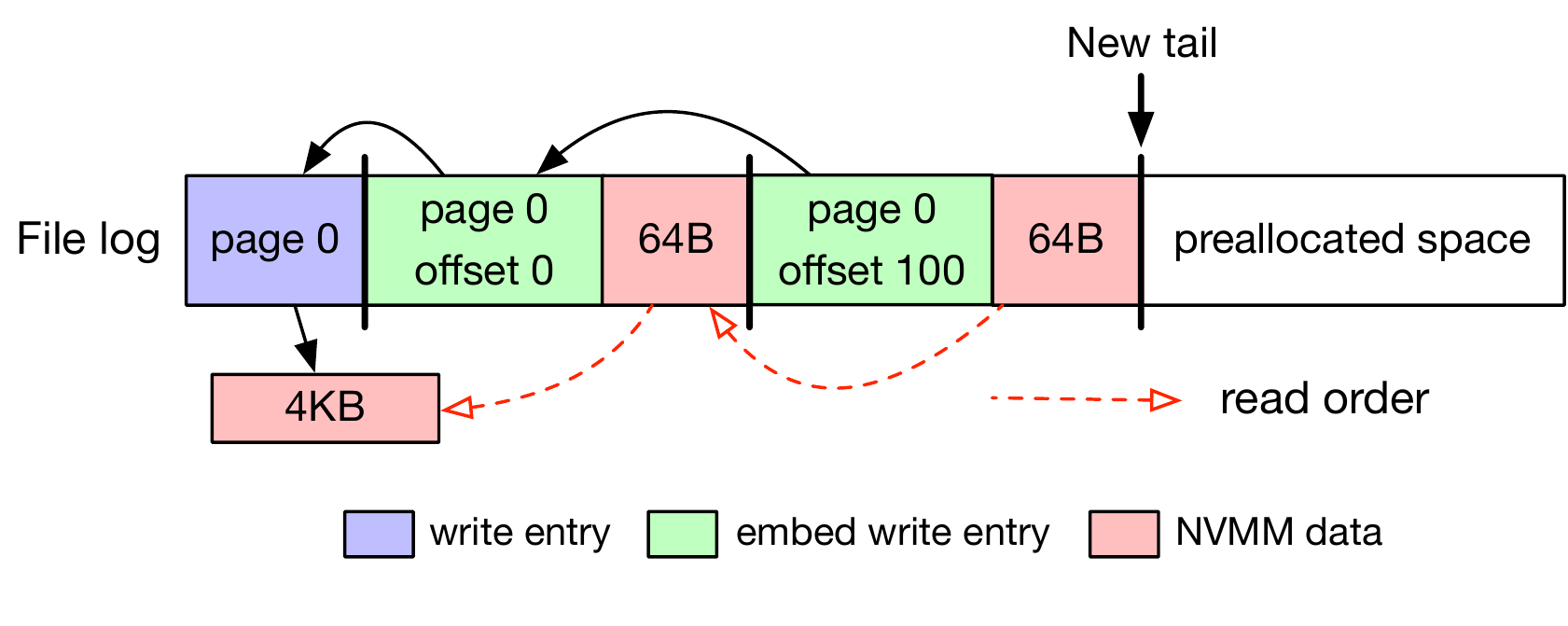,{\figtitle{NOVA-datalog mechanism} Sequentially embedding
data along with metadata turns random writes into sequential writes so that it improves small random write
speed without giving up atomic file updates.
This figure illustrates how NOVA-datalog appends two 64~B random writes
(at 0 and 100, respectively) into the log of a 4~KB file.
},fig:nova-embed-design]

\ntwfigure[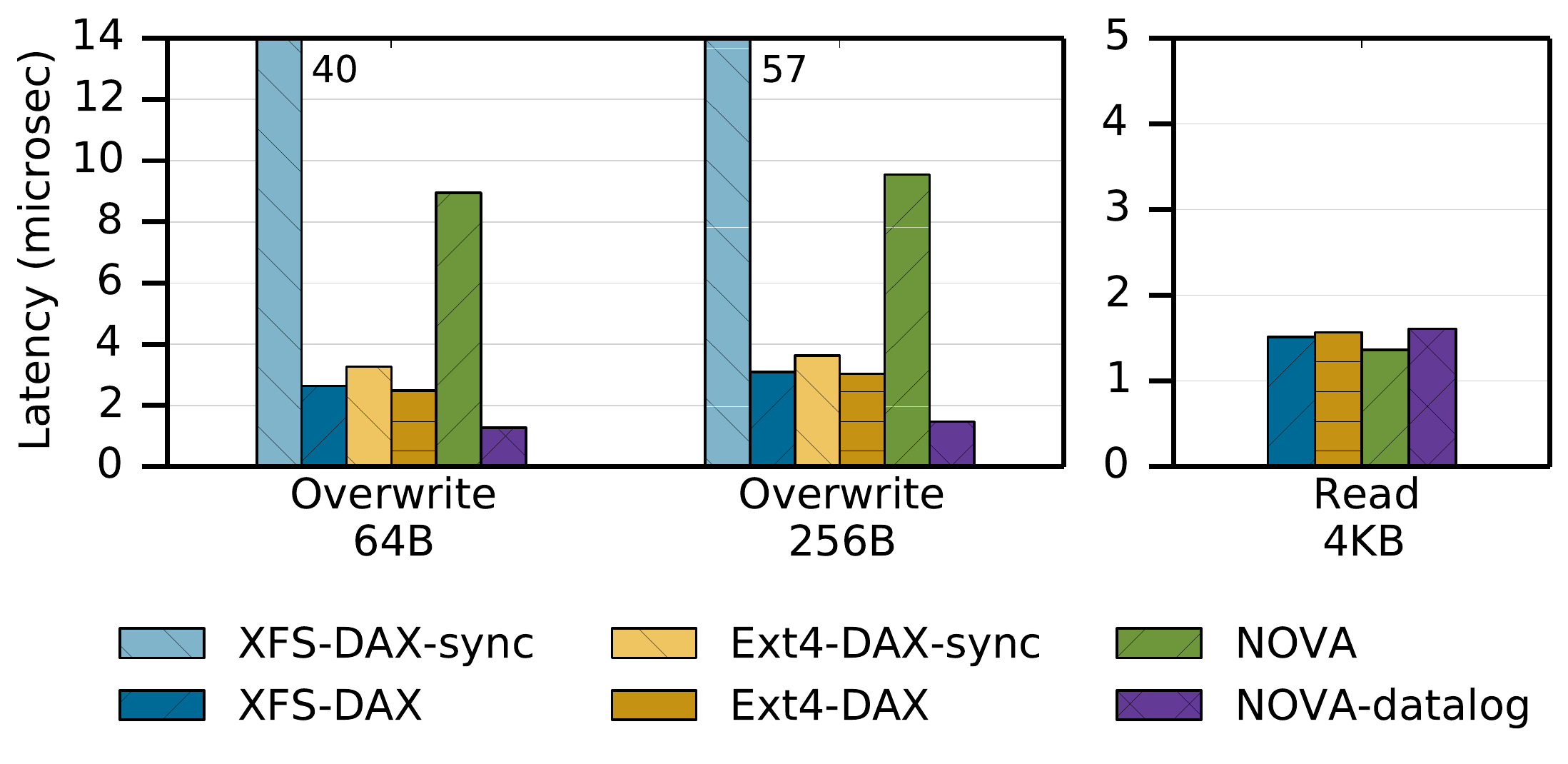,{\figtitle{File IO latency} 
NOVA-datalog significantly speeds up small random writes, but adds a slight overhead in the read path.
Like NOVA and unlike Ext4 or XFS, NOVA-datalog still provides data consistency.
},fig:nova-embed-lat]

The original NOVA design has two characteristics that degrade performance
on \XP{}.  First, the log entries NOVA appends for each metadata update are
small -- 40-64~B, and since NOVA uses many logs, log updates
exhibit little locality, especially when the file system is under load.
Second, NOVA uses copy-on-write to 4~KB pages for file data updates, resulting
in useless stores.  This inefficiency occurs regardless of the underlying
memory technology, but \XP{}'s poor store performance exacerbates its effect.

We address both problems by increasing the size of log entries and avoiding
some copy-on-write operations.  Our modified version of NOVA
-- \emph{NOVA-datalog} -- embeds write data for sub-page writes into the log (\reffig{fig:nova-embed-design}).
Unlike the log's normal \emph{write entry}, which contains a pointer to a new copy-on-write
4~KB page and its offset
within the file, an \emph{embed write entry} contains a page offset, an address within the page, and is followed
by the actual contents of the write. 

This optimization requires several subsidiary changes to the original NOVA
design.  In particular, NOVA must merge sub-page updates into the target page
before memory-mapping or reading the file.  Small changes are also required to the log
cleaner to track the liveness of embedded file data.

\reffig{fig:nova-embed-lat} shows the latencies of random overwrites and reads for three file systems
with different modes.  For XFS-DAX and Ext4-DAX (the two NVM-based file systems included in Linux), we measure both normal \texttt{write} and \texttt{write} followed by \texttt{fsync} (labeled with ``-sync'').

NOVA-datalog improves write performance significantly compared to the original
design (by 7\x{}, 6.5\x{} for 64~byte, 256~byte writes, respectively) and 
meets (for 256~B) or exceeds (for 64~B) performance
for the other file systems (which do not provide data consistency).
Read latency increases slightly compared to the original NOVA.

\ewr{} measurements generally mirror the performance gains.  NOVA-datalog provides the
largest gains relative to NOVA because by merging the data into the log it
eliminates the discontinuity between writes into the log entry and writes into the file
data.

%% file: guideline-ntstore.tex
\subsection{Use non-temporal stores for large writes}

The choice of how programs perform and order updates to \XP{} has a large impact on performance.
When writing to persistent memory, programmers have several options.  After a
regular \pcode{store}, programmers can either evict (\pcode{clflush},
\pcode{clflushopt}) or write back (\pcode{clwb}) the cache line to move the data
into the ADR and eventually into the \XP{} DIMM.  Alternatively, the
\pcode{ntstore} instruction can write directly to persistent memory, bypassing
the cache hierarchy.  For all these instructions, a subsequent \pcode{sfence}
ensures that the effects of prior evictions, write backs, and non-temporal
stores are persistent.

In \reffig{fig:bw-instruction}, we compare achieved bandwidth (left) and latency (right) for sequential accesses
using AVX-512 store instructions with three different instruction sequences: \pcode{ntstore},
\pcode{store + clwb}, and \pcode{store}, followed by a \pcode{sfence}.  Our bandwidth
test uses six threads as it gives good results for all instructions.

The data show that flushing after each 64~B store improves the bandwidth for
accesses larger than 64~B.  We believe this is because letting the cache
naturally evict cache lines adds nondeterminism to the access stream that reaches
the \XP{} DIMM.  Proactively cleaning the cache ensures that accesses remain
sequential.  The \ewr{} correlates this hypothesis: Adding flush
instructions increases the \ewr{} from 0.26 to 0.98.

The data also show that non-temporal stores have lower latency than \pcode{store + clwb}
for accesses over 512~B. Non-temporal stores also have highest bandwidth for accesses over
256~B. Here, we suspect the performance boost is due to the fact that a \pcode{store + clwb}
must load the cache line into the CPU's local cache before executing store, thereby
using up some of the \XPDIMM{}s bandwidth.  As \pcode{ntstore}s bypass the cache,
they will avoid this extraneous read and can achieve higher bandwidth.

In \reffig{fig:flush}, we show how \pcode{sfence}s affect performance.
We use a single thread to issue sequential writes of different sizes on \PMLPMEMUI{}.
We issue \pcode{clwb} during the write of each cache line (every 64B),
or after the entire write (write size). 
At the end of the write we issue a single \pcode{sfence} to ensure the entire
write is persistent (we call this entire operaton an ``sfence interval'').
The result shows the bandwidth peaks
when the write size is 256~B.  This peak is a consequence
of the semantics of \pcode{clflushopt} which is tuned for 
moderately sized writes~\cite{inteloptmanual}.
Flushing during or after a medium sized write (beyond 1~KB) does not affect the bandwidth,
but when the write size is over 8~MB, flushing after the write causes
performance degradation as we incur cache capacity invalidations and a higher \ewr{}.

\ntwfigure[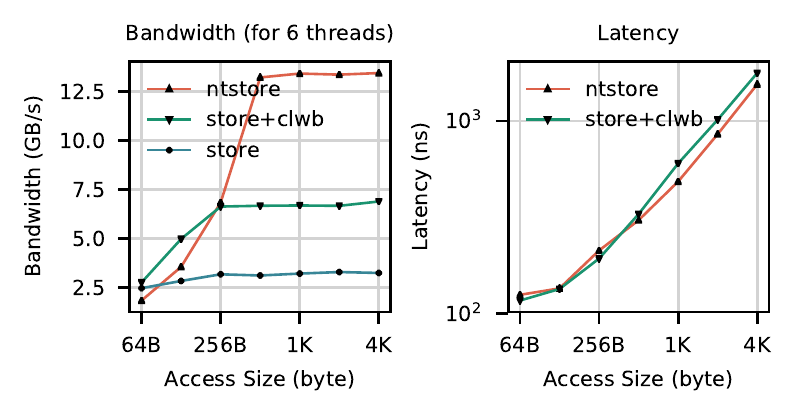,{\figtitle{Performance achievable with persistence instructions} 
Flush instructions have lower latency for small accesses, but \pcode{ntstore} has better latency
for larger accesses.  Using
\pcode{ntstore} also avoids an additional read of the cache line from \XP{} memory,
resulting in higher bandwidth.},fig:bw-instruction]

\ntwfigure[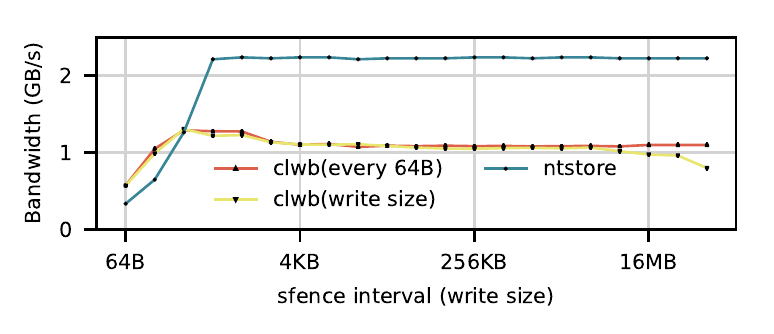,{\figtitle{Bandwidth over \pcode{sfence} intervals.} 
The bandwidth of \XP{} memory decreases when \pcode{sfence} interval increases and when
the window of \pcode{clwb} causing implicit cache evictions.},fig:flush]

	\ntwfigure[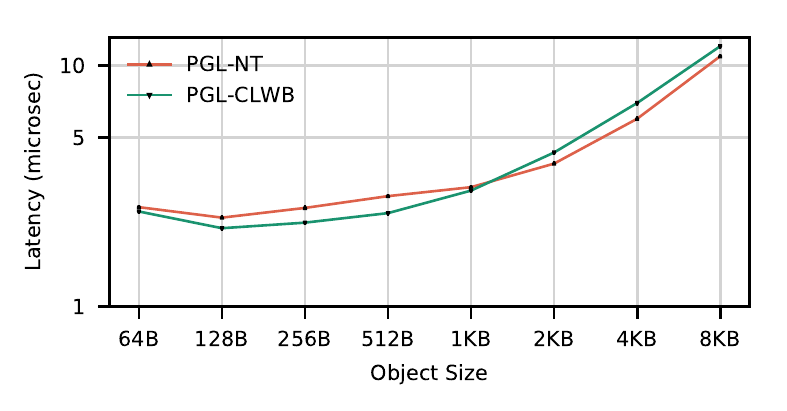,{\figtitle{Tuning persistence instructions for micro-buffering} Using \pcode{ntstores}
only for large writes, and using \pcode{clwb} for small stores can improve performance even at the macro-level.},fig:pangolin-ewr]

\subsubsection{Case Study: Micro-buffering for PMDK}

Our analysis of the relative merits of non-temporal versus normal stores
provides us an opportunity to optimize existing work.  For example, recent work proposed
the ``micro-buffering''~\cite{pangolin} technique for transactionally updating persistent objects.
That work modified the Intel's PMDK~\cite{intel-pmdk} transactional persistent
object library to copy objects from \XP{} to DRAM at the start of a transaction
rather than issuing loads directly to \XP{}.  On transaction commit, it writes back the entire
object at once using non-temporal stores.  

The original paper only used non-temporal stores, but our analysis suggests
micro-buffering would perform better if it used normal stores for small objects
as long as it flushed the affected cache lines immediately after updating them.
\reffig{fig:pangolin-ewr} compares the latency of a no-op transaction for objects of
various sizes for unmodified PMDK and micro-buffering with non-temporal and
normal store-based write back.  The crossover between normal stores and
non-temporal stores for micro-buffering occurs at 1~KB.

%% file: guideline-par.tex
\subsection{Limit the number of concurrent threads accessing a \XP{} DIMM}
\label{sec:interleaving}

Systems should minimize the number of concurrent threads targeting a single
DIMM simultaneously.  An \XP{} DIMM's limited store performance and limited buffering at the iMC and
on the DIMMs combine to limit its ability to handle accesses from multiple
threads simultaneously.  We have identified two distinct mechanisms that
contribute to this effect.

\boldparagraph{Contention in the \RMW{}} Contention for space in the \RMW{}
will lead to increased evictions and write backs to \XP{} media, which will drive
down \ewr{}.  Figure~\ref{fig:bwthreads}(center) shows this effect in action: For
example, when using a single thread to issue sequential non-temporal stores, we
can achieve an \ewr{} of 0.98.  In contrast, when using 8 threads to issue
sequential non-temporal stores to private regions of the DIMM, the \ewr{} drops
to 0.62.

\boldparagraph{Contention in the iMC}
\reffig{fig:align} illustrates how limited queue capacity in the iMC also hurts performance when multiple cores target a single DIMM.
The figure shows an experiment that uses a fixed number of threads (24 for read
and 6 for ntstore) to read/write data to
6 interleaved \XP{} DIMMs. We let each thread access $N$ DIMM (with even distribution
across threads) randomly.  As $N$ rises, the number of writers targeting each
DIMM grows, but the per-DIMM bandwidth drops. A possible culprit is the limited
capacity of the \RMW{}, but \ewr{} remains very close to 1, so the performance
problem must be in the iMC.

On our platform, the WPQ buffer in the iMC cannot queue data more than 256~B
from a single thread.  Our
hypothesis is that, since \XP{} DIMMs are slow, they drain the WPQ slowly,
which leads to head-of-line blocking effects.  Increasing $N$ increases
contention for the DIMMs and the likelihood that any given processor will block
waiting for stores ahead of it to complete.

\reffig{fig:bw}(right) shows another example of this phenomenon --- \XP{} bandwidth falls drastically
when doing random 4~KB accesses across interleaved \XP{} DIMMs. 
\XP{} memory interleaving is similar to RAID-0 in disk arrays:  
The chunk size is 4~KB and the stripe size is 24~KB (Across
the 6 DIMMs on the socket, each gets a 4~KB contiguous block).  The workload in \reffig{fig:bw}(right) spreads accesses across
these interleaved DIMMs, and will lead to spikes in contention for particular DIMMs.

Thread starvation occurs more often as the access size grows, reaching maximum
degradation at the interleaving size (4~KB).
For accesses larger than the interleaving size, each core starts spreading their accesses across multiple DIMMs, evening out the load.
The write data also show small peaks at 24~KB and 48~KB where accesses are perfectly distributed
across the six DIMMs.

This degradation effect will occur whenever 4~KB accesses are distributed
non-uniformly across the DIMMs.  Unfortunately, this is probably a common case
in practice.  For instance, a page buffer with 4~KB pages would probably perform
poorly in this configuration.

\ntwfigure[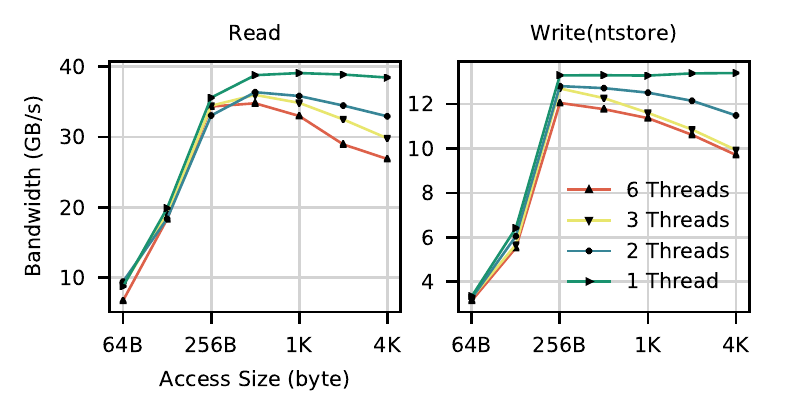,{\figtitle{Plotting iMC contention.} 
With a fixed number of threads (6), as the number of DIMMs accessed by each thread grows, bandwidth
drops.  For maximal bandwidth, threads should be pinned to DIMMs.},fig:align]

\subsubsection{Case Study: Multi-NVDIMM NOVA}

The original NOVA design did not attempt to limit the number of writers per
DIMM.  In fact, as, it tends to allocate pages for a file from contiguous regions
which, via interleaving, tends to spread
those pages across the DIMMs.
To fix this issue, we configured our machine to pin writer threads to non-interleaved
\XP{} DIMMs.  This configuration ensures an even matching between threads and NVDIMMs, thereby leveling
the load and maximizing bandwidth at each NVDIMM.

\ntwfigure[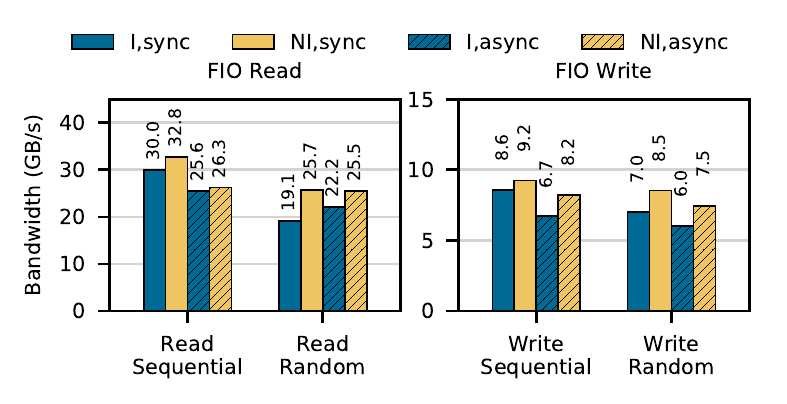,{\figtitle{Multi-DIMM NOVA} Evenly
distributing the load across NVDIMMs is important for maximizing
application bandwidth. By making NOVA multi-DIMM aware, we improve performance
by an average of 17\%
on the FIO benchmark.
},fig:multi-dimm-nova]

\reffig{fig:multi-dimm-nova} shows the result.
Our experiment uses the FIO
benchmark~\cite{fio} to test the optimization and uses 24 threads.
We plot the bandwidth of each file operation with two 
different IO engines: \pcode{sync} and \pcode{libaio} (async).
By being Multi-NVDIMM aware, our optimization improves NOVA's bandwidth by between 3 and
34\%.

%% file: guideline-numa.tex
\subsection{Avoid mixed or multi-threaded accesses to remote NUMA nodes}

NUMA effects for \XP{} are much larger than they are for DRAM, so designers
should work even harder to avoid cross-socket memory traffic.  The cost is
especially steep for accesses that mix load and stores and include multiple
threads.  Between local and remote \XP{} memory,
the typical read latency difference is 1.79\x{} (sequential) and 1.20\x{} (random),
respectively. For writes, remote \XP{} memory's latency is 2.53\x{} (ntstore)
and 1.68\x{} higher compared to local. For bandwidth, remote \XP{}
can achieve 59.2\%{} and 61.7\%{} of local read and write bandwidth
at optimal thread count (16 for local read, 10 for remote read and 4 for local and remote write).

The performance degradation ratio above is similar to remote DRAM to local DRAM.
However, the bandwidth of \XP{} memory is drastically degraded
when either the thread count increases or the workload is read/write mixed.
Based on the results from our systematic sweep, the bandwidth gap between
local and remote \XP{} memory for the same workload can be over 30\x{},
while the gap between local and remote DRAM is, at max, only 3.3\x{}.

In \reffig{fig:numa}, we show how the bandwidth changes for \XP{} on both local and
remote CPUs by adjusting the read and write ratio.  We show the performance of
a single thread and four threads, as local \XP{} memory performance increases
with thread count up to four threads for all the access patterns tested.

Single-threaded bandwidth is similar for local and remote accesses.  For
multi-threaded accesses, remote performance drops off much more quickly as store
intensity rises, leading much lower performance relative to the local case.

\ntwfigure[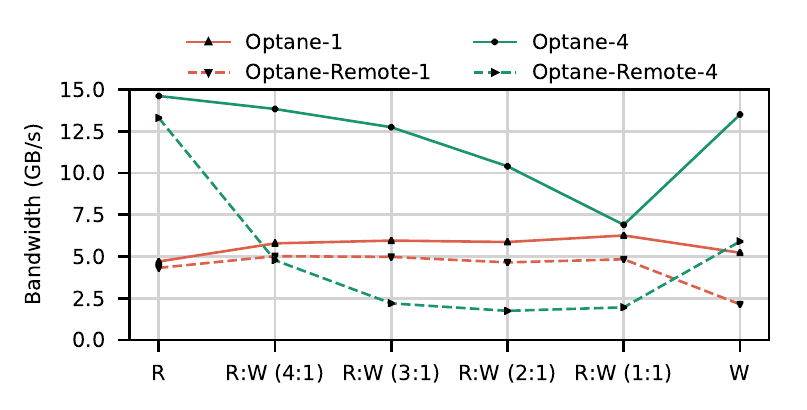,{\figtitle{Memory bandwidth on \PMLPMEM{}
and \PMRPMEM{}} This chart shows memory bandwidth as we vary the mix of accesses
for both one and four threads.  Pure reads or pure writes perform better on NUMA than mixed
workloads and increased thread count generally hurts NUMA performance.},fig:numa]

\ntwfigure[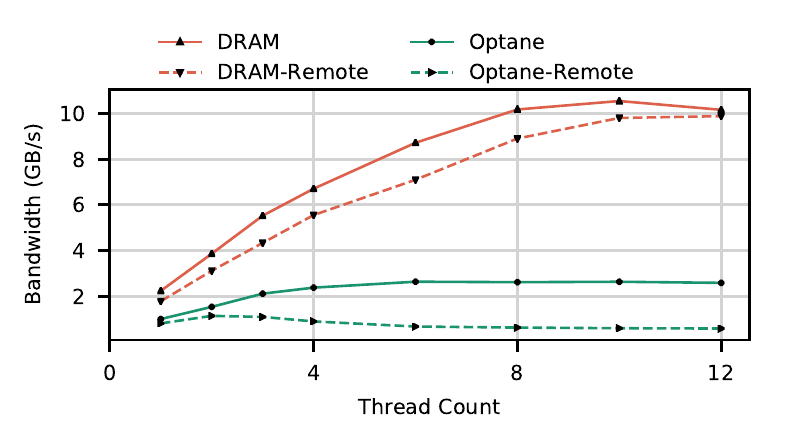,{\figtitle{NUMA degradation for PmemKV} \XP{} memory
experiences greater NUMA-based degradation than DRAM.  In our experiments, migrating
to a remote \XP{} node reduces the performance of the persistent key-value store PmemKV by 
up to 4.5\x{} (18\x{} versus DRAM)},fig:pmemkv-numa]

\subsubsection{Case Study: PMemKV}

Intel's Persistent Memory Key-Value Store (PMemKV~\cite{intel-pmemkv}) is an NVMM-optimized 
key-value data-store. It implements various index data structures (called ``storage engines'') 
to index programs data and uses the Persistent Memory Development Kit (PMDK~\cite{intel-pmdk}) 
to manage its persistent data. We used the concurrent hash map (\pcode{cmap}) in our
tests as it is the only engine that supports concurrency.

To test the effect of \XP{}'s NUMA imbalance on PMemKV, we varied the location of
the server relative to the \pcode{pmem} pool; Figure~\ref{fig:pmemkv-numa} shows the result
on the included benchmark on
a test with mixed workload (\pcode{overwrite}) that repeatedly performs read-modify-write operations.
In this test, using a remote \XPDIMM{} drops the performance of the store beyond two threads.
\XP{} performance is far more drastically impacted by the migration (loss of ~75\%) 
than DRAM (loss of ~8\%).

%% file: limitations.tex
\section{Discussion}
\label{sec:limitations}

The guidelines in \refsec{sec:best_practice} provide a starting point for
building and tuning \XP{}-based systems.  By necessity, they reflect the
idiosyncrasies of a particular implementation of a particular persistent memory
technology, and it is natural to question how applicable the guidelines will be
to both other memory technologies and future versions of Intel's \XP{} memory.
It is also important to note that we have only studied the guidelines in the context of App Direct mode,
since the large DRAM cache that Memory Mode provides mitigates most or all of the effects they account for.
We believe that our guidelines will remain valuable both as \XP{} evolves and
as other persistent memories come to market.

The broadest contribution of our analysis and the resulting guidelines is that
they provide a road map to potential performance problems that might arise in
future persistent memories and the systems that use them.  Our analysis shows
how and why issues like interleaving, buffering on and off the DIMM, and
instruction choice, concurrency, and cross-core interference can affect
performance.  If future technologies are not subject the precisely the same
performance pathologies as \XP{}, they may be subject to similar ones.

Ultimately it is unclear how commercially available, scalable persistent memories
will evolve.  Several of our guidelines are the direct product of
(micro)architectural characteristics of the current \XP{} incarnation.  The
size of the \RMW{} and iMC's WPQ might change in future implementations, which
would limit the importance of minimizing concurrent threads and reduce the
importance of the 256~B write granularity.  However, expanding these structures
would increase the energy reserves required to drain the ADR during a power
failure.  Despite this, there are proposals to extend the ADR down to the
last-level cache~\cite{WSP,kiln} which would eliminate the problem.
An even more energy-intensive change would be to make the DRAM cache
that \XP{} uses in Memory mode persistent.

Increasing or decreasing the 256~B internal write size is likely to be expensive.
It is widely believed that \XP{} is phase-change memory and the small internal page
size has long been a hallmark of the phase change memory~\cite{onyx} due to power
limitations.  Smaller internal page sizes are unlikely because the
resulting memories are less dense.

A different underlying memory cell technology (e.g., spin-torque MRAM) would
change things more drastically.  Indeed, battery-backed DRAM is a well-known
and widely deployed (although not very scalable or cost-effective) persistent
memory technology.  For it, most of our guidelines are no longer needed, 
although non-temporal stores are still more efficient for
large transfers due to the restrictions of the cache coherency model.

%% file: related.tex
\section{Related Work}
\label{sec:related}

With the release of the \XPDIMM{} in April 2019, early results on the devices
have begun to be published. 
For instance, Van Renan
et al.~\cite{alex2019persistent} have explored logging mechanisms for the devices,
and Izraelevitz et al.~\cite{izraelevitz2019basic} 
have explored general performance characteristics.  
We expect additional results to be published in the near future as the devices
become more widely available.

Prior art in persistent memory programming has spanned the
system stack, though until very recently these results
were untested on real \XP{} media.  
A large body of work has explored 
transactional memory-type abstractions for enforcing a consistent persistent
state~\cite{mnemosyne, nvheaps, atlas,
justdologging, phytm,idologging,nvthreads,giles-msst-2015,
kaminotx}.
Various authors have built intricate NVM data structures
for data storage and transaction processing~\cite{oukid-sigmod-2016,
yang-fast-2015, chen-vldb-2015, 
chatzistergiou-vldb-2015,
dalinvm, izraelevitz-disc-2016,
lockfreenvmqueue,bztree,pmemcached,cdds}. 
Recent research on in-memory databases has also investigated NVM-based
durability.
For online transactional processing (OLTP)
engines not explicitly designed for NVM,
NVM-aware logging~\cite{huang-vldb-2014, wang-vldb-2014,
fang-icde-2011,pelley-vldb-2014} and query processing~\cite{viglas-vldb-2014} can
improve performance.  Other authors have investigated speculative techniques
for adapting database designs for architectures with NVM~\cite{debrabant-vldb-2014, arulraj-sigmod-2015},
while Kimura's FOEDUS~\cite{foedus} builds a custom DBMS
for NVM from the ground up.  Custom NVM file systems have 
also been explored~\cite{novapaper,orion,octopus,scmfs,aerie,BPFS,pmfs,nova-fortis,ziggurat},
as have more speculative hardware architectures~\cite{vorpal,ISCA2018,pmorder,whisper}.

%% file: conclude.tex
\section{Conclusion}
\label{sec:conclude}

This paper has described the performance of Intel's new \XPDIMM{}s
across micro- and macro-level benchmarks.  In doing so, we have extracted
actionable guidelines for programmers to fully utilize these devices strengths.
As expected, the devices have performance characteristics that lie in-between
traditional storage and memory devices, yet they also present a number of interesting
performance pathologies and pitfalls that the programmer must be careful to avoid.
That said, we believe that the devices will be useful in extending the quantity
of memory available to memory-intensive applications and in providing low-latency storage
to high performance applications.  Future work remains in migrating existing systems
to these devices and designing custom software for their peculiarities.

%% file: bib.tex
\bibliography{libpaper/common,paper}
\bibliographystyle{plain}